\documentclass[
    ,final            
    ,numberedheadings 
    ,cmfonts          
  ]
  {aipproc}
\input epsf
\usepackage{epsfig}
\usepackage{fancyhdr}
\usepackage{latexsym}
\usepackage{amsmath}
\usepackage{amssymb}
\usepackage[latin1]{inputenc}
\usepackage{layout}
\layoutstyle{6x9}
\newcommand{\jpsi}{{\ensuremath{J/\psi}}}
\newcommand{\ups}{{\ensuremath{\Upsilon}}}

\newcommand{\ptjpsi}{{\ensuremath{\rm P_{T}^{J/\psi}}}}

\newcommand{\ptups}{{\ensuremath{\rm P_{T}^{\Upsilon}}}}

\newcommand{\ccbar}{\ensuremath{{\rm c\bar{c}}}}
\newcommand{\bbbar}{\ensuremath{{\rm b\bar{b}}}}

\newcommand{\ptaa}{{\ensuremath{\rm P_{T}^{R=0.7}}}}
\newcommand{\pthat}{{\ensuremath{\rm \hat{p}_{T}}}}

\begin{document}

\title{Experimental Aspects of Heavy Quarkonium Production at the LHC}

\classification{14.40.Gx, 13.85.Ni, 25.75.Dw}
\keywords      {Quarkonium, Production, LHC experiments}

\author{A.C.~Kraan\footnote{Financed by EU as Marie Curie Research Fellow}
}{
  address={Istituto Nazionale di Fisica Nucleare, Largo Pontecorvo 3, 56100, Pisa, Italy\\
E-mail: Aafke.Kraan@cern.ch}
}

\begin{abstract}
More than 30 years after the discovery of the $J/\psi$, its production mechanism is still poorly understood. With the LHC data it will be possible to study quarkonia up to very high transverse momenta and with high statistics. In this note we discuss experimental aspects of \jpsi~and \ups(1S)~production at the LHC. In particular, we investigate the sensitivity of a general purpose LHC detector to observables, which are complementary to the cross section and polarisation measurement.
These observables would be sensitive to the radiation produced in association with the quarkonium. 

\end{abstract}

\maketitle


\section{Introduction}
Although \jpsi's and \ups's have been studied extensively since their discoveries, their prompt production mechanism is still not well understood. For reviews on quarkonium production the reader is referred to Refs~\cite{yellow,lansberg}. Below we only summarise some relevant aspects. 

Initially, prompt production  of \jpsi's and \ups's was assumed to take place via the leading order \emph{colour singlet model} (CSM).
However, in 1997, CDF studies showed large excesses in the amount of prompt quarkonia produced with respect to theory predictions~\cite{Abe:1997jz, Abe:1997yz}, and new production mechanisms were invented. Among the most popular approaches is Non-Relativistic QCD, where production at parton level can also take place via a colour octet quark pair. While this \emph{colour octet mechanism} (COM) is quite successful in explaining the transverse momentum spectra of quarkonia, the polarisation predictions of the COM mechanism are in sheer disagreement with experimental data: production via a colour-octet state predicts a significant transverse polarisation of \jpsi'~s and {\ups's}, while recent Tevatron data contradict this prediction~\cite{cdfpol_jpsi, d0pol_ups}. The CSM in its own has been revived recently, when it has been shown that NLO and NNLO QCD contributions~\cite{maltoni,lansberg1,lansberg2} as well as $s$-channel cut contributions~\cite{CSM-NNLO,Lansberg:2005pc} arising from relativistic corrections could lead to an enhancement of the production cross section of the CSM,
 while predicting a polarisation closer to the data. 

Given the puzzling situation, it is the utmost importance to study prompt quarkonia at the LHC, where they can be studied with vast statistics up to high transverse momentum. Standard measurements are generally related to kinematical distributions of the quarkonium decay products, such as differential cross section and polarization measurements, and focus on decays into muons. Although these provide useful information, they may be sensitive to other factors. Hence it is important to investigate additional observables. In particular, we should take into account not only the kinematics of the quarkonium itself, but also that of particles produced in association. In fact, in some cases prompt quarkonia can be searched for in association with a heavy quark pair~\cite{lansberg2}. Also, for an unambiguous understanding of quarkonium production, direct production should be studied, for example $\chi_c$ production could be studied by radiative decays into \jpsi's.

In this note we discuss general aspects of prompt \jpsi~and \ups(1S) production at the LHC. 
Also, we show examples of observables that are sensitive to the hadronic activity directly around the quarkonium, thereby allowing to extract information about the radiation emitted off the coloured heavy quark pair during production. The results apply only to a general purpose LHC detector, and for this work we took the CMS parameters~\cite{cmstdr} as reference. We do not distinguish here between direct and indirect (via radiative $\chi_c$ decay) prompt \jpsi-production.

The outline of this note is as follows. In Section 2 we discuss event generation of \jpsi~and \ups(1S) in PYTHIA. In Section 3 we discuss the production of \jpsi's at the LHC, and study the differences in radiation between different production toy models. We will also study the \jpsi-reconstruction efficiency. In Section 4 we concentrate on \ups's. Section 5 contains conclusions.

\section{Quarkonium event generation}
We have produced \jpsi-events using PYTHIA 6.409~\cite{pyt}. The implementation~\cite{tormar} is based on the NQRCD approach and quarkonium production via the singlet (LO) and octet mechanism. The octet \ccbar-states have a mass of 3.1 GeV. The values of the NRQCD matrix elements are taken from Ref.~\cite{marianne}. We have modified the default branching ratios for the radiative decays of the $\chi$-states into \jpsi's to correspond to the recent PDG values~\cite{pdg}. The polarisation has been set to zero. Production via $\chi$ states is included. In this study we forced the \jpsi~and \ups~to always decay into $\mu^+\mu^-$, the branching ratios are 5.93\% and 2.48\% respectively~\cite{pdg}. The following features are included.
\subsection{Cross section regularisation} We have made use of an event reweighting function, allowing the quarkonia cross section to be dampened at small $P_T$ values, in analogy to the underlying event formalism in PYTHIA~\cite{zijl, ps}. The events obtain a weight $w_i$ according to
\begin{equation}\label{eq:wi}
w_i=\frac{\sigma_{reweighted}}{\sigma_{not\hspace{0.9mm} reweighted}}=\Big(\frac{\hat{p}_T^2}{p^{2}_{T_0}+\hat{p}_T^2}\Big)^2\Big(\frac{\alpha_S(p^{2}_{T_0}+Q^2)}{\alpha_S(Q^2)}\Big)^3
\end{equation}
where $\hat{p}_T$ is the transverse momentum evaluated in the rest frame of the scattering, $Q^2$ the momentum transfer scale, and $p_{T_0}$ the regularisation scale, used also in the underlying event cross section regularisation. As can be seen from Eq.~\ref{eq:wi}, at small \ptjpsi a suppression is obtained while at large \ptjpsi the cross section is unmodified. The value of $P_{T_0}$ is be extrapolated from the $P_{T0}$ value determined at 1960 GeV to the LHC energy:
\begin{equation}\label{eq:pt0}
p_{T_0}=PARP(82)\Big( \frac{E_{cm}}{PARP(89)}\Big)^{PARP(90)} = 1.94\rm{GeV}\Big( \frac{14 \rm{TeV}}{1.96\rm{TeV}}\Big)^{0.16}\approx 2.66~\rm{GeV}
\end{equation}
In Fig.~\ref{xs} (left) we demonstrate the regularisation procedure, as well as the difference between the old PYTHIA (before version 6.324) and the new production model.

\subsection{Shower activity associated with quarkonium production}\label{pytshower}
Below we discuss \jpsi-production, however it applies to \ups's too. If the \jpsi~is produced via the singlet model, the \ccbar~does not radiate and the charmonium is thus produced in isolation (when produced at LO), up to normal underlying event activity. When it is produced via the octet mechanism, the situation is not apriori clear, but it is reasonable to assume that a shower can evolve and that the charmonium is embedded in a certain amount of shower activity. This results from 1) repeated $g\rightarrow gg$ branchings before the branching $g\rightarrow \ccbar$, and 2) from the radiation off the colour octet \ccbar-state once formed. 
There are several switches  to regulate the hadronic activity. Using these, we generate \jpsi's in four different production mechanisms, each with varying amount of radiation.

\begin{itemize}
\item Production via \emph{singlet} only.  The \jpsi's are produced in isolation. In the hard process a number of hard gluons is emitted for colour neutralization, but these are recoiling against the \jpsi~and thus expected to be back-to-back to the \jpsi.
\item Octet production only, where the octet $c\bar{c}$ undergoes only a small amount of parton shower evolution only. Here MSTP(148)=0, implying that the splitting kernel used for the branching $\ccbar^{(8)}\rightarrow\ccbar^{(8)}g$ is the $q\rightarrow qg$ kernel, modified for the \ccbar-mass. Only a small amount of activity if expected  to be emitted during the parton cascade, yet the activity is expected to be much larger than in the colour singlet case. This case is referred to as \emph{octet low radiation} scenario.
\item Octet production only, but MSTP(148)=1 and MSTP(149)=0. Now the splitting kernel for $\ccbar^{(8)}\rightarrow\ccbar^{(8)}g$ is the $g\rightarrow gg$ one, again modified for the \ccbar-mass. In each branching the \ccbar-state is likely to get the largest energy: $z>0.5$. The amount of radiation is larger than in the previous case, however because of the suppression of $z<0.5$ only a small amount of hard gluons is expected around the \jpsi. This case is referred to as \emph{octet medium radiation}.
\item Octet production only, but now MSTP(148)=1 and MSTP(149)=1. Again the $g\rightarrow gg$ splitting kernel is used, but now $z>0.5$ and $z<0.5$ have equal probability (in reality $z>0.5$ still dominates due to mass effects). Here a significant amount of hard gluons can be emitted before the hadronisation process, in fact showering activity might be overestimated. We will refer to this case as \emph{octet high radiation} scenario.
\end{itemize}
It must be emphasised that these models are 'straw-man models', and that we do not expect any of them to reproduce the data by itself. 
Finally, note that CDF data can be reproduced by adding singlet and octet contributions, but that any of the 3 octet models together with the singlet model fit the data. 
\subsection{Event generation parameters}\label{genpar}
For the production of prompt \jpsi$\rightarrow\mu^+\mu^-$-events we have generated events in 5 bins of $\hat{p}_T$, the transverse momentum exchanged in the hard process (0-10, 10-20, 20-30, 30-50, 50-$\inf$ GeV). This was done in view of the large production cross section at low \ptjpsi, in order to have statistics also at large \ptjpsi. In total 2M events in four different production scenarios (see Sec.~\ref{pytshower}) have been generated. The luminosity of the lowest \pthat-bin for the singlet model corresponds to only 0.2 pb$^{-1}$,while the higher \pthat bins correspond to higher luminosities. We applied a filter at generator level requiring two muons with $|\eta|<2.5$ and $p_T>2$ GeV/$c$.

As background for the \jpsi-production studies we have considered non-prompt \jpsi-events. An inclusive non-prompt \jpsi~sample has been generated (PYTHIA setting MSEL=1, cross section 55 mb), and passed through a dimuon generator level filter ($|\eta|<2.5$ and $p_T>2.5$ GeV/$c$). The luminosity of the sample we use is 4.45 pb$^{-1}$, corresponding to roughly 400000 filtered non-prompt J/psi-events.

Prompt \ups's have been generated in the four different production scenarios of Sec.~\ref{pytshower}, corresponding to 100 pb$^{-1}$,  without \pthat-bins. Here a dimuon filter was used requiring the muons to have $|\eta|<2.5$ and $p_T>3$ GeV/$c$. Since the \ups-production cross section is smaller than that of \jpsi's, we will need more LHC data, possibly including pile-up. We included here 5 pile-up events per \ups-event.

As background for the \ups-production studies we have used an inclusive $pp\rightarrow\mu X$ sample generated again with MSEL=1 in PYTHIA, corresponding to 0.14 pb$^{-1}$.

All events are processed through a toy-detector based on the expected performances of the CMS detector~\cite{cmstdr}. We have not included trigger requirements, but being interested in high values of the transverse momentum of the quarkonium (see Sec 4), we expect no problems in triggering. Typically, a double muon with some threshold would be used, for example 3 GeV for each muon~\cite{cmstdr}.
\section{\jpsi~production studies}
\subsection{\jpsi~production at the LHC}
\begin{figure}[t!]
\epsfig{file=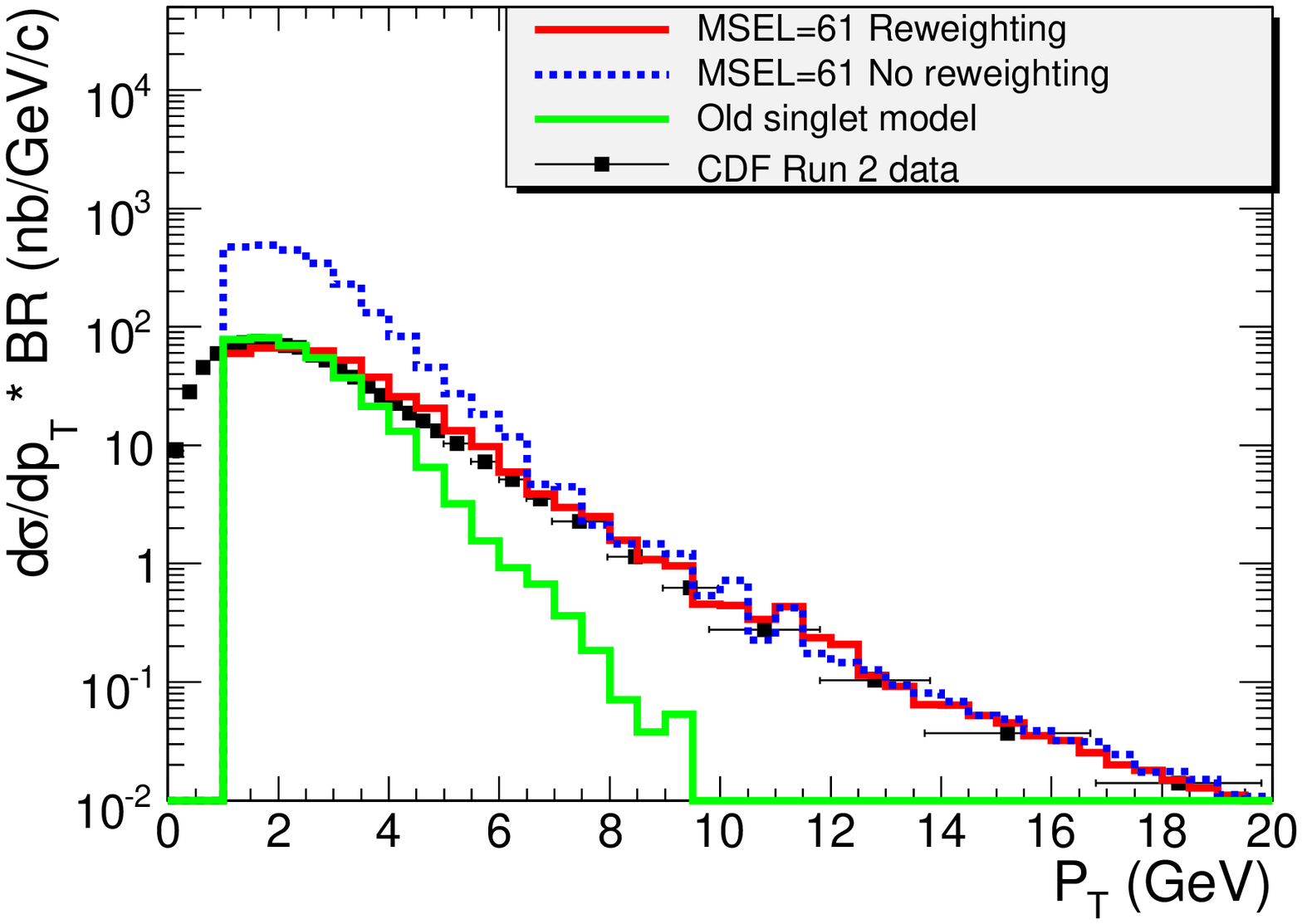,width=7.5cm}\epsfig{file=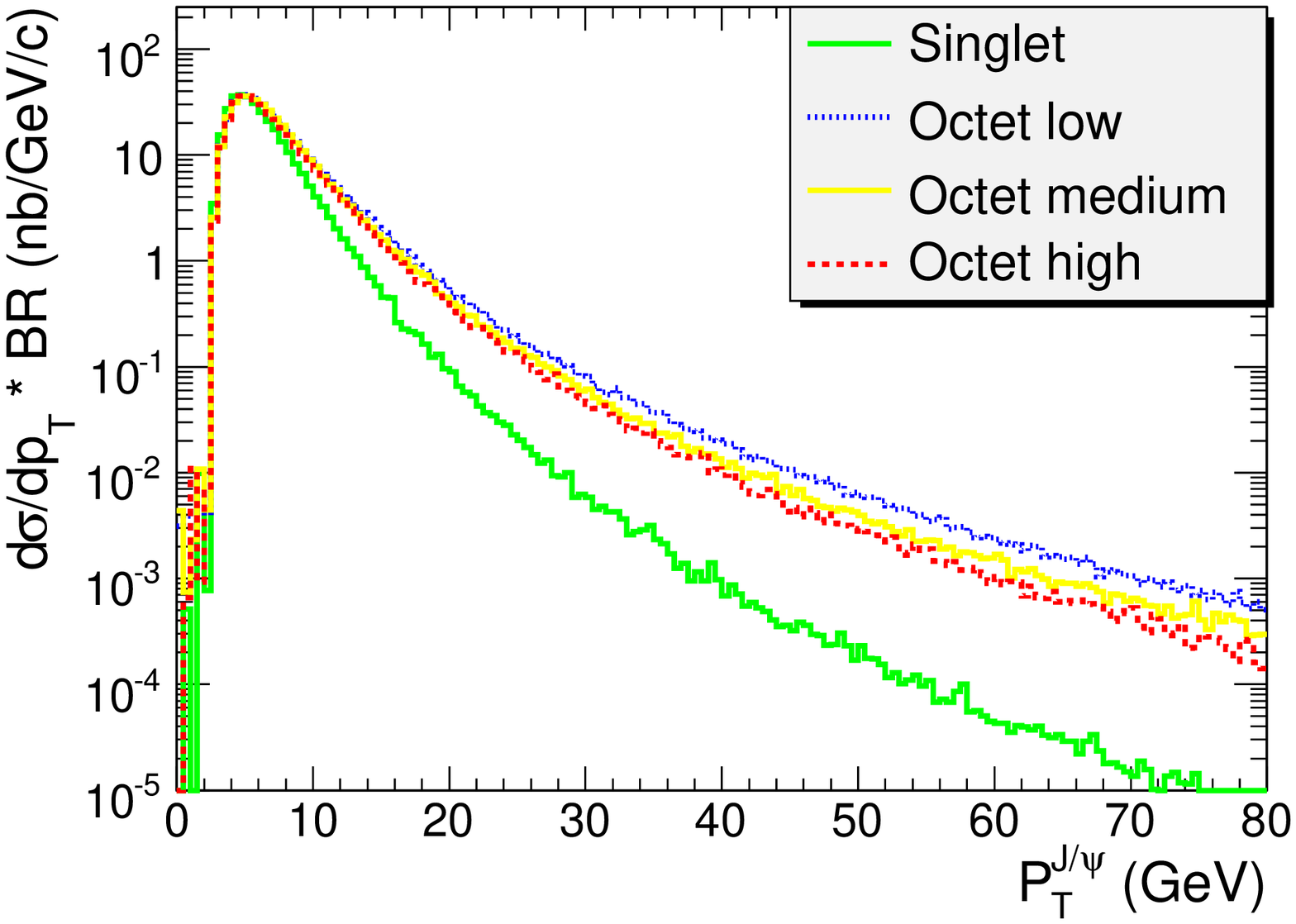,width=7.5cm}
\caption{\footnotesize{Left: \jpsi's at the Tevatron. The old model (singlet only, normalised to fit CDF data, green line), the new model without reweighting (blue broken line), the new model with reweighting (red line), and the CDF data (black data points, from Ref.~\cite{cdfxs_jpsi}). Right: differential cross section at CMS-like detector for the four production cases separately. \label{xs}}}
\end{figure} 

In Fig.~\ref{xs} (right) the differential cross section for a CMS-like detector as function of \ptjpsi~is shown for the four production cases.
To obtain the total cross section the singlet curve should be added with one of the three octet curves. The singlet curve decreases rapidly with \ptjpsi, as expected. For the octet cases, the decrease is faster for models with more radiation, since here the \ccbar~loses more energy in radiating.
Although the curves are well separated at high transverse momentum, these curves are influenced by different factors,
such as initial and final state radiation, the mass of the cc-octet, and parameters entering in the cross section dampening~\cite{talkhamburg}.
In conclusion, it is important to find complementary observables.

To have a rough idea of the statistics expected at CMS, we display in Tab.~\ref{tab:numbers} the numbers of \jpsi's produced within acceptance in ranges of \ptjpsi~in 100 pb$^{-1}$.
\begin{table}[b!]
\begin{tabular}{|l|c|c|}
\hline
\ptjpsi-range (GeV) & Nr. produced & Nr. reconstructed\\
\hline
0-10 & $3.0\times 10^7$ & $3.7\times 10^6$\\
10-20& $3.5\times 10^6$ & $1.6\times 10^6$\\
20-30& $2.0\times 10^5$ & $1.4\times 10^5$\\
30-40& $3.2\times 10^4$ & $2.2\times 10^4$\\
40-50& $7.8\times 10^3$ & $5.6\times 10^3$\\
$>$50& $3.5\times 10^3$ & $2.4\times 10^3$\\
\hline
\end{tabular}
\caption{\footnotesize{Rough estimate for number of prompt \jpsi's produced and reconstructed in 100 pb$^{-1}$ (singlet and octet medium radiation summed). Reconstruction of \jpsi's is discussed in Sec.~\ref{sec:rec}. \label{tab:numbers}}}
\end{table} 

\begin{figure}[t!]
\vspace*{-0.1cm}
\epsfig{file=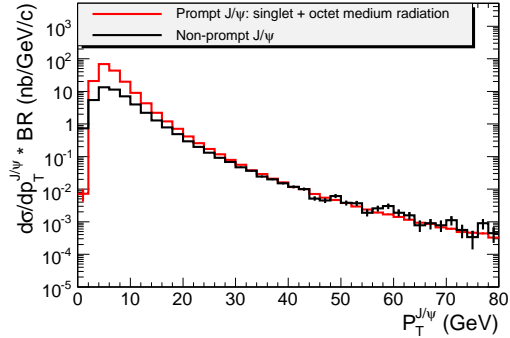,width=7.0cm}
\caption{\footnotesize{Differential cross section for non-prompt (black) and prompt (red) \jpsi's.\label{fig:xsnonprompt}}}
\end{figure}

Concerning the production of non-prompt \jpsi's, the rate at LHC is significant, as can be seen in Fig.~\ref{fig:xsnonprompt}. While at low \ptjpsi~prompt production dominates, at higher \ptjpsi~the non-prompt and prompt contributions are comparable. 

\subsection{\jpsi~reconstruction}\label{sec:rec}
\begin{figure}[htbp]
\hspace*{0cm}\epsfig{file=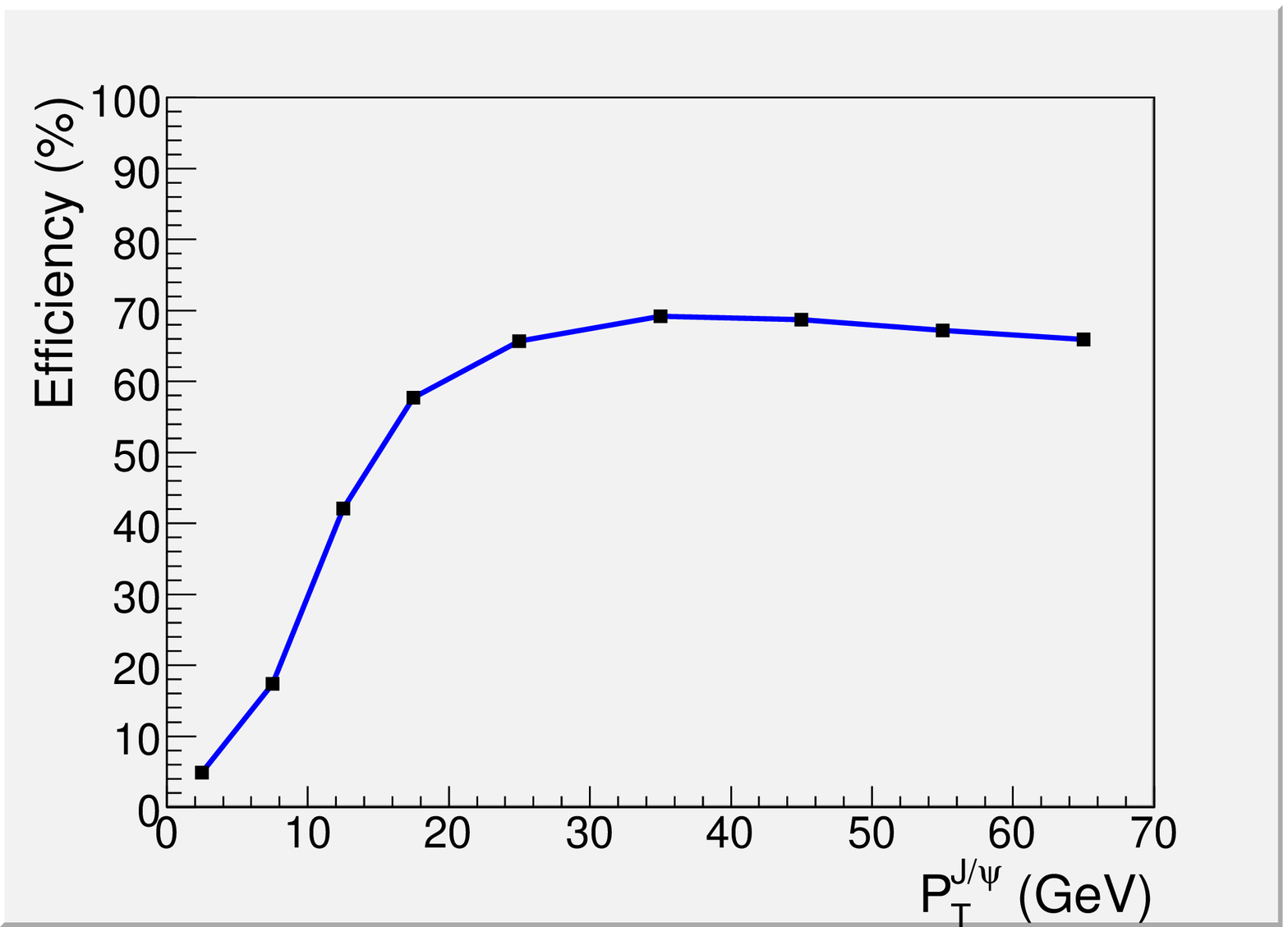,width=5.2cm}\hspace{0.9cm}\epsfig{file=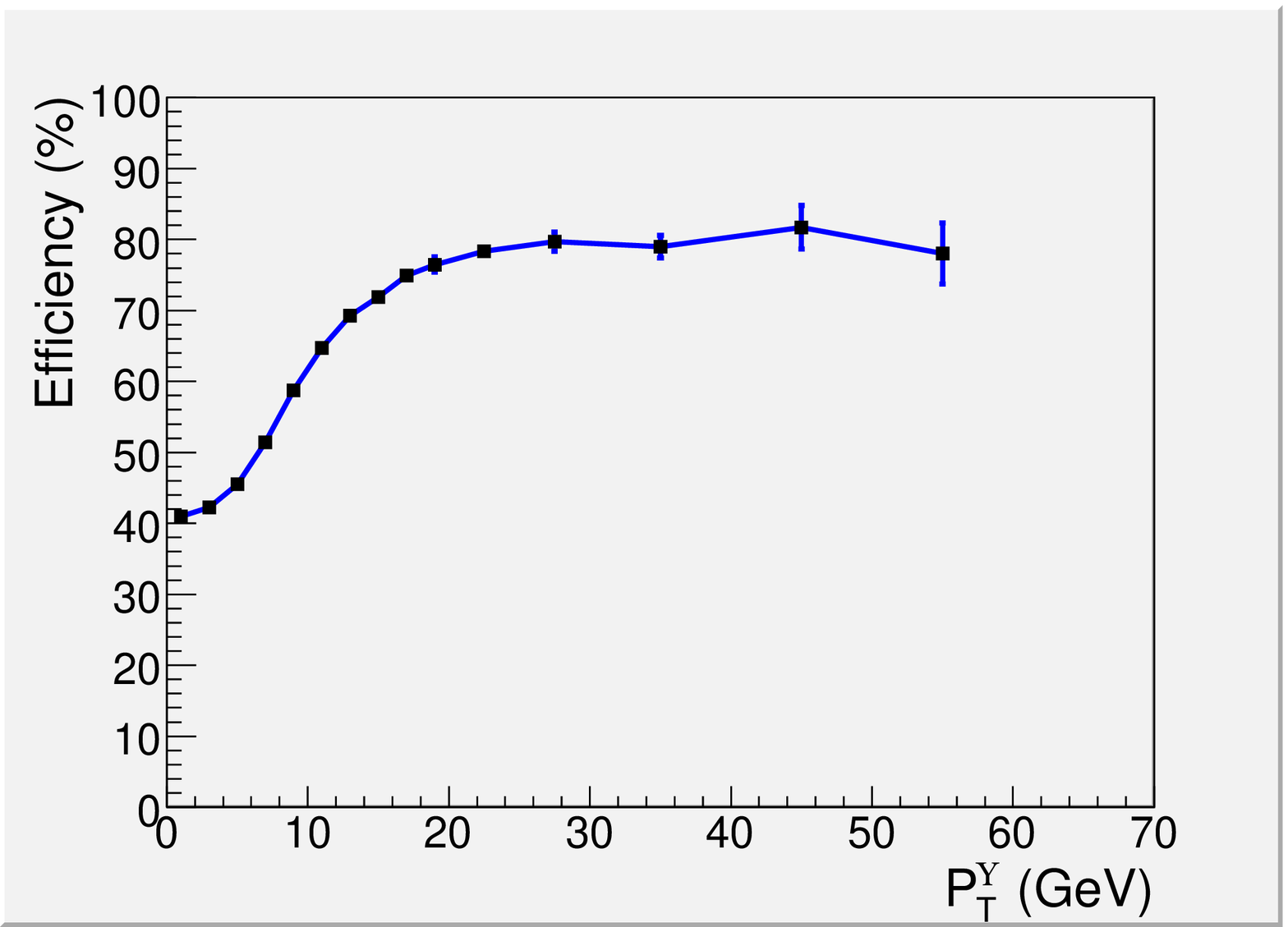,width=5.2cm}
\caption{\footnotesize Left: expected \jpsi~reconstruction efficiency in general purpose LHC detector as function of \ptjpsi. Right: the same but for \ups's. (see Sec.~\ref{upsrec})\label{zyeff}}
\end{figure}

Reconstruction of \jpsi-events with the CMS detector is discussed in Ref.~\cite{zongchang}. We display in Fig.~\ref{zyeff} the reconstruction efficiency with respect to the filtered events of our toy detector as function of \ptjpsi. As long as no isolation requirements are applied, the reconstruction efficiency is independent of the production model. It can be seen that \jpsi's with $P_T$ below 3 GeV are hardly reconstructable in a detector like ATLAS or CMS, because these \jpsi's decay into two approximately back-to-back muons with too low momentum.
It must be noted that both ALICE and LHCb are able to  reconstruct \jpsi's with much lower transverse momentum. 
\subsection{Sensitivity to alternative \jpsi-production observables}
Since the shower-activity is different for the different models, we investigated several variables sensitive to the kinematics of the particles produced in association with the \jpsi. Because at lower \ptjpsi, shower activity is small in general, differences manifest themselves only at higher values of \ptjpsi, about 20 GeV/c. It must be noted that the energy of the surrounding particles associated with \jpsi-production is small (order GeV), and as such it is not apriori clear whether there is any sensitivity at all. 

\begin{figure}[h!]
\epsfig{file=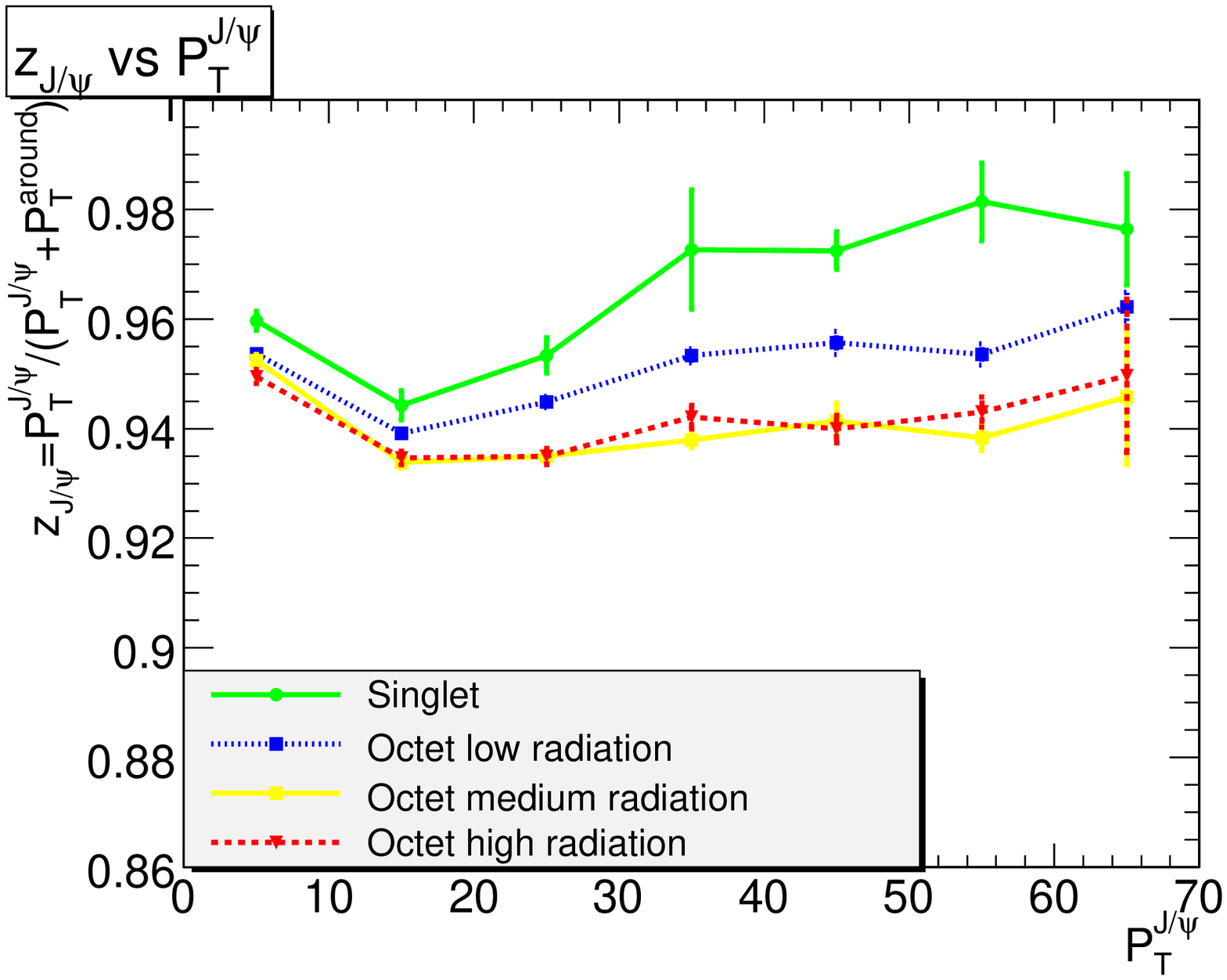, width=6cm}\hspace{0.9cm}\epsfig{file=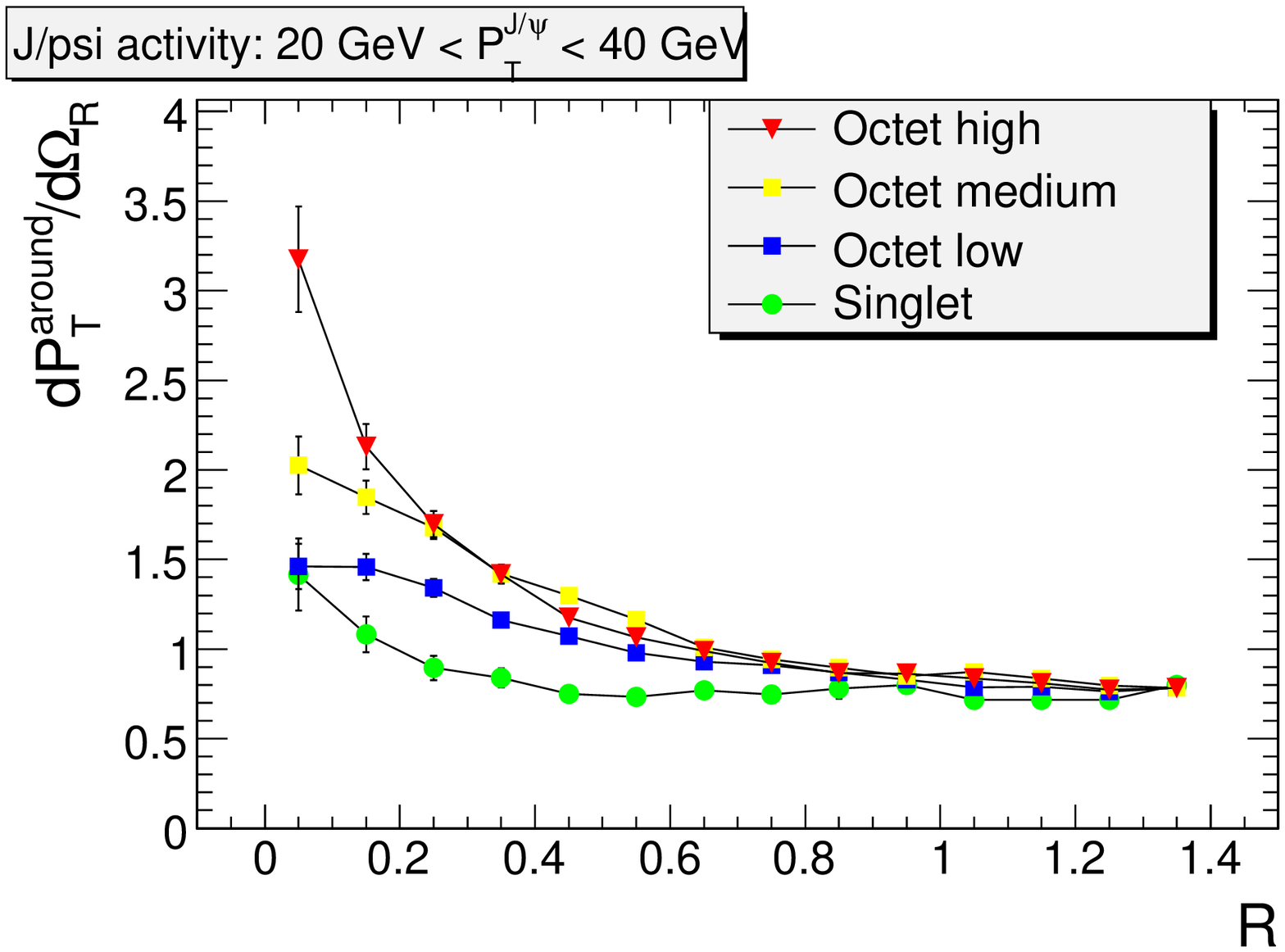,width=6cm}
\caption{\footnotesize Left: variable $z_{\jpsi}$ at reconstruction level. Right: variable $\frac{\rm d P_{T}^{around}(R)}{\rm d\Omega_R}$ at reconstruction level. \label{z}}
\end{figure}

Examples of observables were shown in Ref.~\cite{hamburg}. In Fig.~\ref{z} (left) we display for the four prompt J/psi production models the fragmentation variable $z_{\jpsi}={\ptjpsi}/{(\ptjpsi+\ptaa)}$ as function of \ptjpsi, after reconstruction criteria. Here \ptaa~is the sum of the transverse momentum of particles in a cone of $R=0.7$. We remark that $z$ is influenced quite strongly by the underlying event activity. In fact, without underlying event activity $z\approx 1$ for the singlet case.

In Fig.~\ref{z} (right) we display the transverse momentum density d$P_T$/d$\Omega_R$ for \jpsi's between 20 and 40 GeV after reconstruction, in a cone around the \jpsi~of certain size $R=\sqrt{(\Delta\eta)^2+(\Delta\phi)^2}$, where
\begin{equation}
\frac{\rm d P_{T}^{around}(R)}{\rm d\Omega_R}=\frac{\rm P_{T}^{around}(\rm  R+d R/2)-P_{T}^{around}(R-dR/2)}{\rm \pi[( R+dR/2)^2-(R-dR/2)^2]}.
\end{equation}
Here $\rm P_{T}^{around}(R)$ is the sum of the transverse momentum of all charged particles (with $P_T>0.9$ GeV) inside the cone of size $R$. At this stage, we can conclude that the particles produced in association with prompt \jpsi-production, even though low energetic, are well detectable with a general purpose detector at LHC.   
\subsection{Background}
The main backgrounds for prompt \jpsi-production are non-prompt \jpsi's and fake \jpsi's from QCD background. The latter can most probably be easily subtracted with sideband subtraction, but for non-prompt \jpsi's the situation is tricky. By studying the impact parameter of the reconstructed \jpsi's, and knowing the behaviour of prompt  and non-prompt \jpsi's, it is possible to evaluate the amount of non-prompt background in data. However, to correctly subtract the non-prompt contribution from data, one must know very precisely the hadronic activity around non-prompt \jpsi's. It turns out that the procedure is difficult, in particular because the hadronic activity around non-prompt \jpsi's is very large in comparison with that of prompt \jpsi's. When attempting to subtract the background, the differences between the prompt \jpsi-production models were washed out. Several improvements are possible, such as more Monte Carlo statistics, or by combining different variables, to gain sensitivity. However, given the fact that for \ups's no non-prompt background is present, we first investigate hadronic activity there.
\section{Upsilon production studies}
\subsection{\ups~production at the LHC}

\begin{figure}[t!]
\epsfig{file=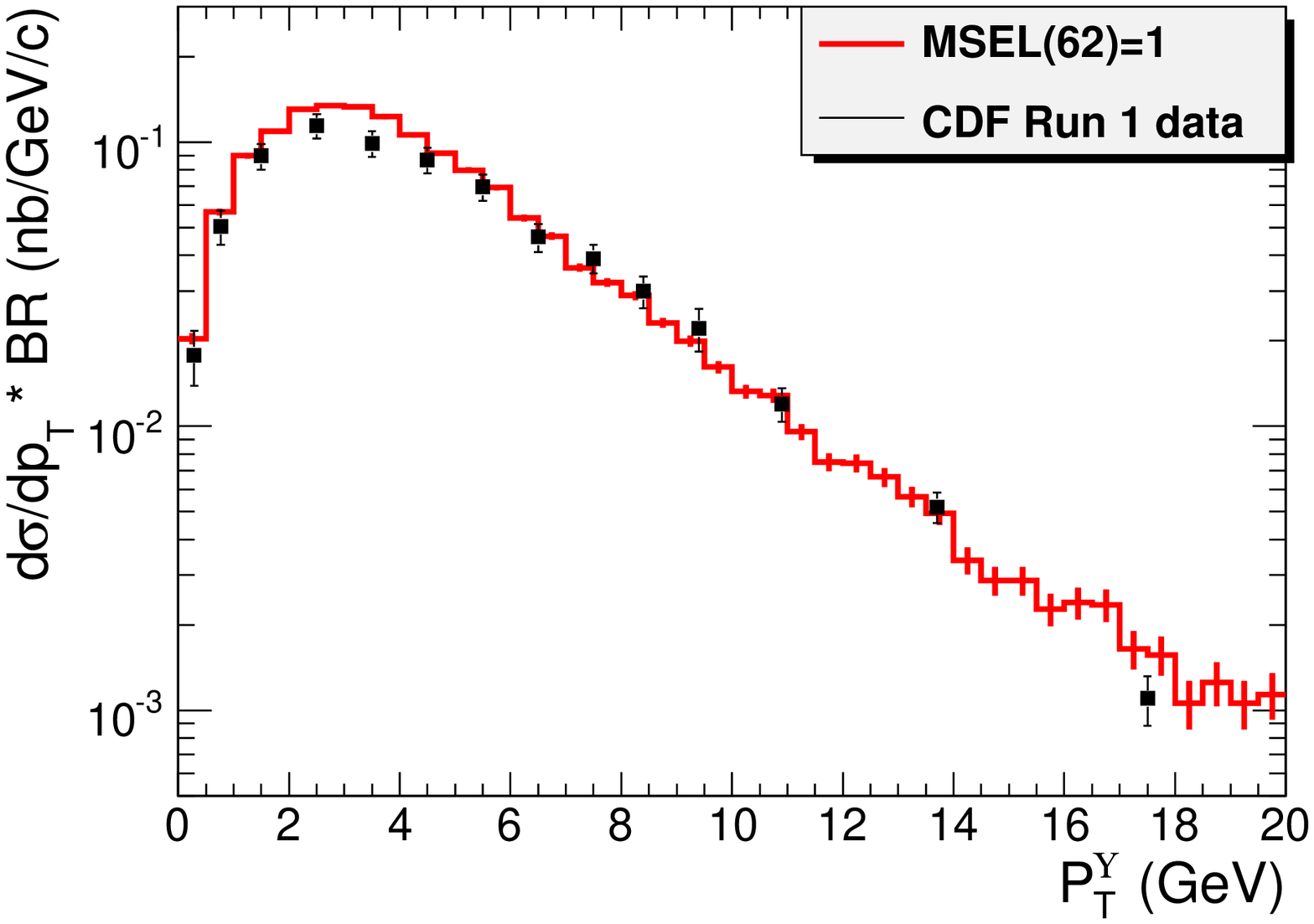,width=6.cm}\epsfig{file=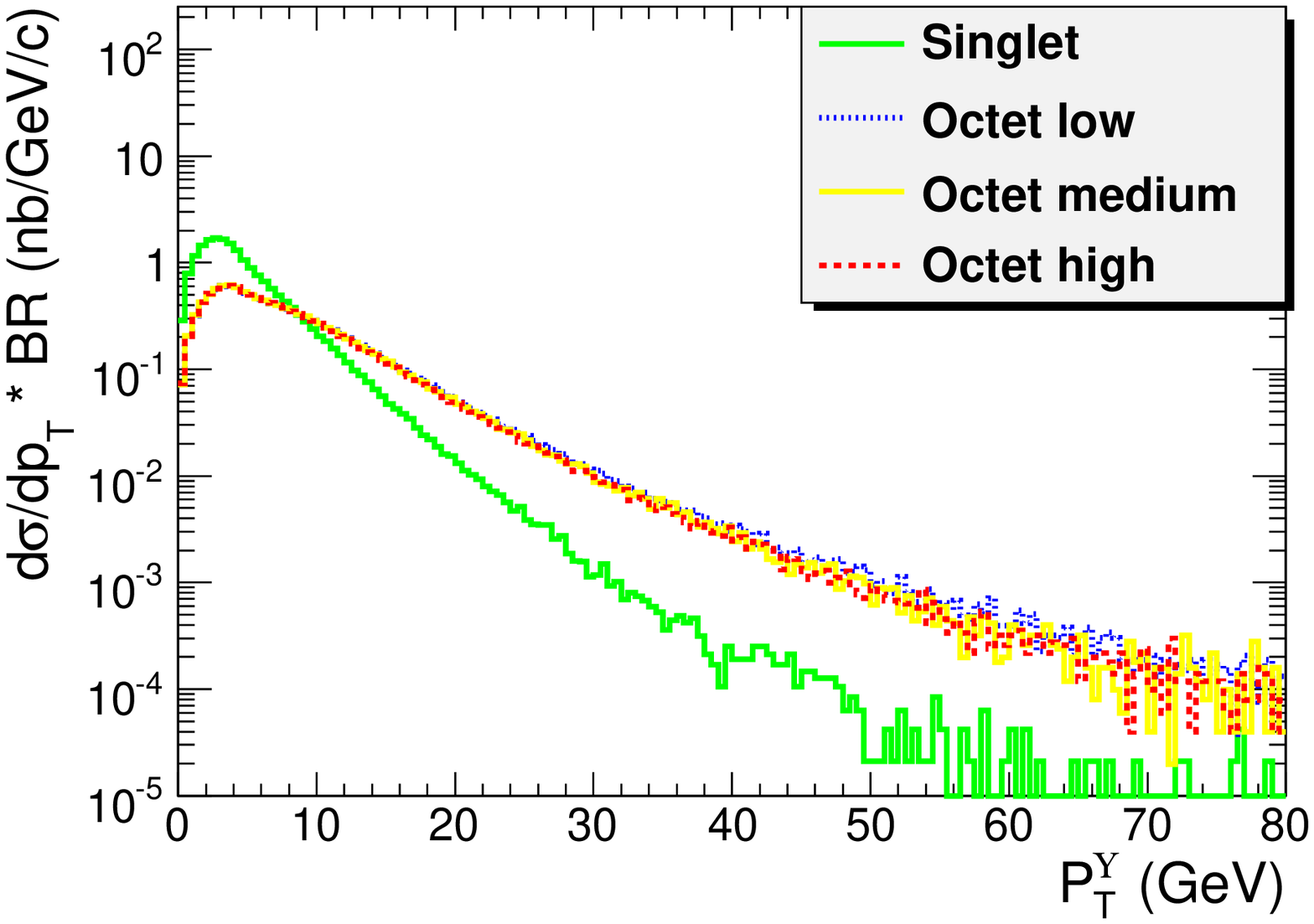,width=6.cm}
\caption{\footnotesize{Left: \ups~production with singlet and octet in PYTHIA (red), together with CDF data (back dots). The latter are taken from Ref.~\cite{cdfxs_ups}. Right: Differential cross section at CMS-like detector for the four production cases separately. \label{xsups}}}
\end{figure}

We generated \ups(1S)-events as described in Sec.~\ref{genpar}. In Fig.~\ref{xsups} (right) the differential cross section for prompt \ups(1S) at LHC as function of \ptups~is shown for the four production cases. In comparison with the \jpsi-cross section in Fig.~\ref{xs}, 
we see that the curves for \ups's fall less rapidly than for \jpsi's. Furthermore, although the difference between the singlet and octet curves is still large, the differences between the three octet models is small, and the curves are almost overlapping.  The reason is that a coloured \bbbar-state emits less gluon radiation than a \ccbar~because it is heavier. 

A rough estimate of the statistics expected at a general purpose detector like CMS or ATLAS is given in Tab.~\ref{tab:numbers2}. 
\begin{table}[t!]
\begin{tabular}{|l|c|c|}
\hline
\ptjpsi-range (GeV) & Nr. produced & Nr. reconstructed\\
\hline
0-10 & $1.3\times 10^6$ & $5.7\times 10^5$\\
10-20& $2.1\times 10^5$ & $1.4\times 10^5$\\
20-30& $3.0\times 10^4$ & $2.3\times 10^4$\\
30-40& $6.1\times 10^3$ & $4.8\times 10^3$\\
40-50& $1.7\times 10^3$ & $1.4\times 10^2$\\
$>$-50& $9.1\times 10^2$ & $7.0\times 10^2$\\

\hline
\end{tabular}
\caption{\footnotesize{Estimate for number of prompt \ups's produced and reconstructed (singlet and medium radiation model added) in 100 pb$^{-1}$. Reconstruction of \ups's is discussed in Sec.~\ref{upsrec}.\label{tab:numbers2}}}
\end{table} 
\subsection{Upsilon reconstruction}\label{upsrec}
Just like \jpsi's a general purpose detector could detect \ups's by detecting the muons. The reconstruction efficiency with respect to the filtered events as a function of \ptups~if displayed in Fig.~\ref{zyeff} (right). As opposed to \jpsi's, \ups's can be reconstructed down to zero transverse momentum. Indeed, even when \ptups$\rightarrow 0$, the momentum of the then back-to-back muons is sufficiency large to be detected, because of the high \ups-mass (9.46 GeV). Furthermore, the value of the reconstruction efficiency where it plateaus is seen to be larger for \ups's than for \jpsi's. This is due to the fact that the muons from $\Upsilon\rightarrow\mu^+\mu^-$ have a larger separation angle than those of $J/\Psi\rightarrow\mu^+\mu^-$, so are better detectable.
\subsection{Sensitivity to alternative observables}
\begin{figure}[h!]
\epsfig{file=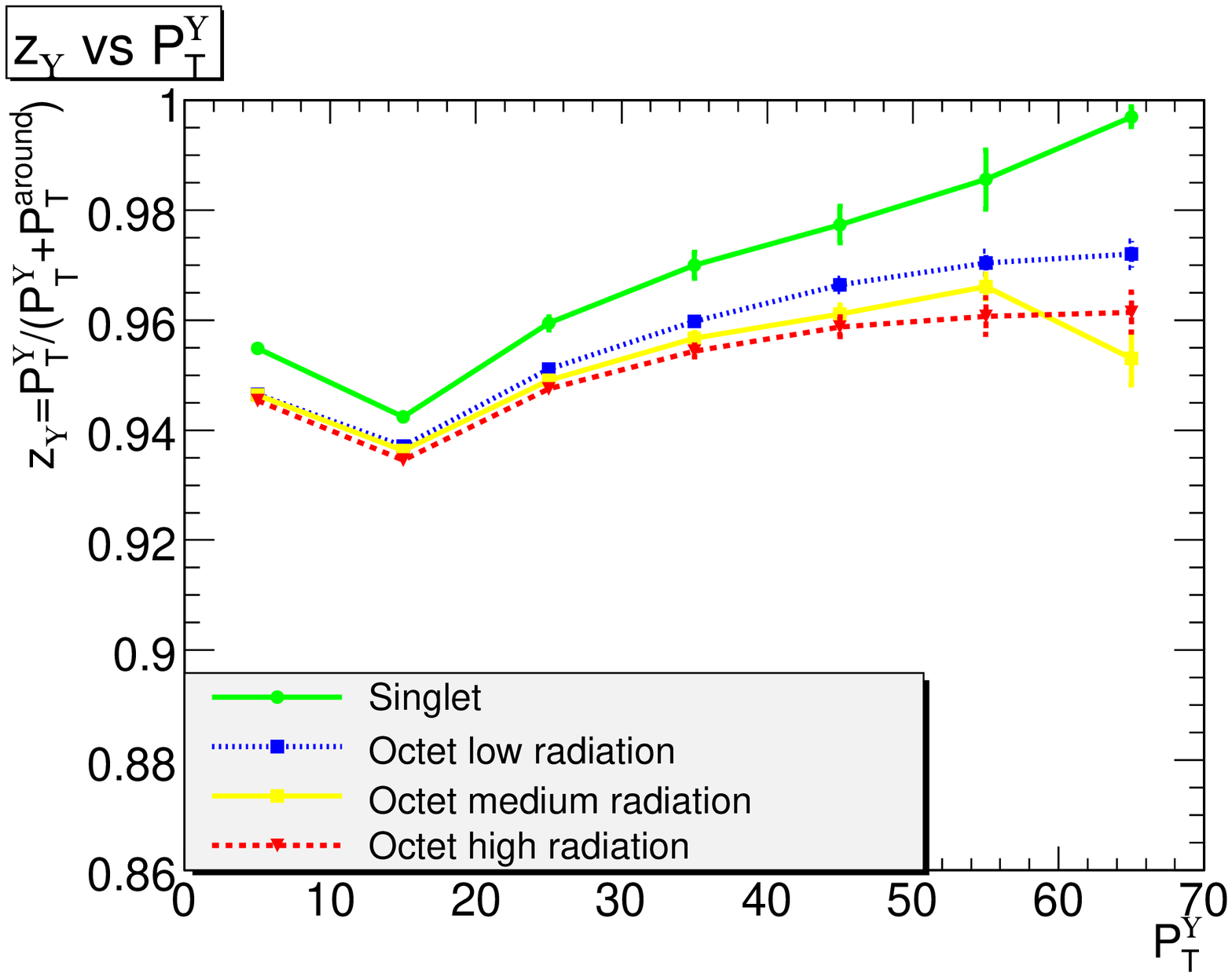, width=6cm}\hspace{0.9cm}\epsfig{file=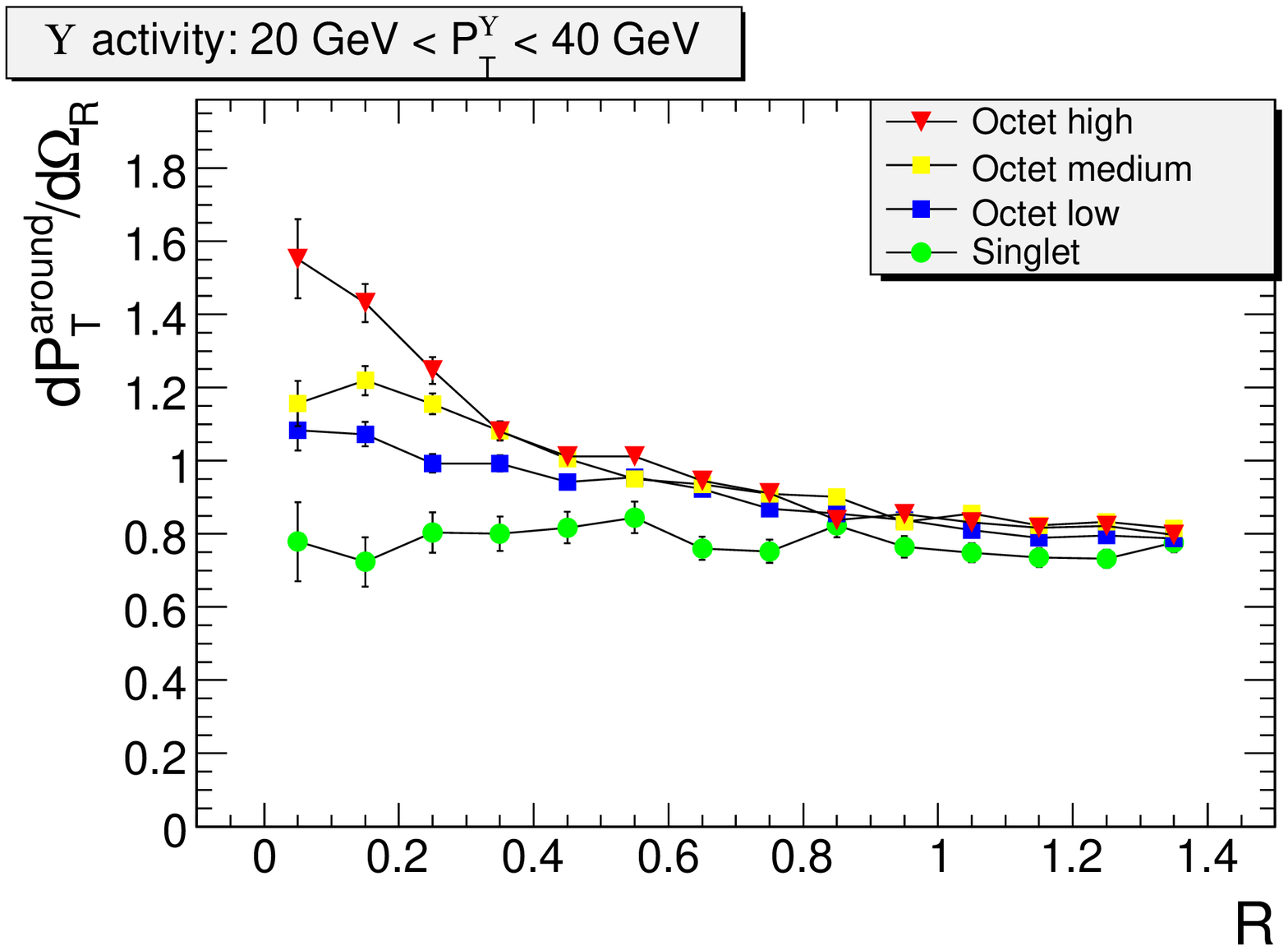,width=6cm}
\caption{\footnotesize Left: variable $z_{\Upsilon}$ at reconstruction level for \ups's. Right: variable $\frac{\rm d P_{T}^{around}(R)}{\rm d\Omega_R}$ at reconstruction level for \ups's. \label{upsact}}
\end{figure}

In Fig.\ref{upsact} (left) we display the variable $z_{\ups}$ at reconstruction level. In Fig.\ref{upsact} (right) the transverse momentum density d$P_T$/d$\Omega_R$ is shown for \ups's between 20 and 40 GeV for the four \ups~production models, in a cone $R$ around the \ups, at reconstruction level. Although the amount of hadronic activity is smaller than for \jpsi's, the differences are still well visible. To obtain these results, first the tracks coming from pile-up have been subtracted.

\subsection{Background}
The main background is from QCD events, where we have b-quarks decaying into muons. 
The spectrum is well-known. Due to the weak b-decay, the hadronic activity around these reconstructed muon-pairs is high. Unfortunately, studying this background with the current Monte Carlo samples is difficult. The efficiency for a $pp\rightarrow\mu X$ background event to end up both close to the \ups-mass and at high \ptups, is very tiny. To study the background with sufficient statistics, a more dedicated Monte Carlo generation should be done. In real data, however, we expect studying background not to be a limiting factor to this analysis, because it can be studied accurately in the sidebands left and right of the \ups-mass window. Having no other background components, most probably it can simply be subtracted to obtain the hadronic activity of pure \ups's. 
\section{Conclusions}
Up to today the prompt quarkonium production mechanism is not understood. LHC will open up new possibilities for studying quarkonium production. Apart from cross section and polarisation measurements, it is important to study different observables. An example is shown here, and suggests studying observables sensitive to hadronic activity produced in association with prompt quarkonia.
\begin{theacknowledgments}
Many thanks to Jean-Philippe Lansberg and the organisation for inviting me to Spa! Thanks to Jean-Philippe also for comments on this note. Furthermore, I thank Torbj\"orn Sj\"ostrand for support and fruitful discussions, and for comments on this note. Many thanks to Zongchang Yang for discussions and help, and thanks to Urs Langenegger, Fabrizio Palla and Carlos Lourenco for useful suggestions.
\end{theacknowledgments}


\begin{thebibliography}{9}
\bibitem{yellow} N.~Brambilla {\it et al}, CERN Yellow Report, CERN-2005-005, e-Print: hep-ph/0412158.
\bibitem{lansberg}
J.~P.~Lansberg,
Int.\ J.\ Mod.\ Phys.\  A {\bf 21} (2006) 3857
\bibitem{Abe:1997jz}
F.~Abe {\it et al.}  [CDF Collaboration],
Phys.\ Rev.\ Lett.\  {\bf 79} (1997) 572.
\bibitem{Abe:1997yz}
F.~Abe {\it et al.}  [CDF Collaboration],
Phys.\ Rev.\ Lett.\  {\bf 79} (1997) 578.
\bibitem{cdfpol_jpsi}
A.~Abulencia {\it et al.}  [CDF Collaboration],
Phys.\ Rev.\ Lett.\  {\bf 99} (2007) 132001
\bibitem{d0pol_ups} D0 Collaboration, Preliminary Result, Conference Note 5089-CONF, July 2007
\bibitem{maltoni} J.~Campbell, F.~Maltoni, F.~Tramontano, Phys.\ Rev.\ Lett.\ {\bf 98} (2007) 252002.
\bibitem{lansberg1} P.~Artoisenet, J.~P.~Lansberg, F.~Maltoni, Phys.\ Lett.\ B {\bf 653} (2007) 60. 
\bibitem{lansberg2}H.~Haberzettl , J.~P.~Lansberg, Phys.\ Rev.\ Lett.\ {\bf 100} (2008) 032006. 
\bibitem{CSM-NNLO} Paper in preparation. See P.~Artoisenet, this volume.
\bibitem{Lansberg:2005pc}
J.~P.~Lansberg, J.~R.~Cudell and Yu.~L.~Kalinovsky,
Phys.\ Lett.\  B {\bf 633} (2006) 301
\bibitem{cmstdr} CMS Technical Design Report Volume 1, CERN/LHCC 2006-001 (2006).
\bibitem{pyt} PYTHIA 6.409: T.~Sj\"ostrand, S.~Mrenna, P.~Skands, JHEP 0605:026,2006. 
\bibitem{tormar} The original implementation
 was done by S.~Wolf in 2000, 
but it was never
included 
officially
In 2005 it was officially integrated by T.~Sj\"ostrand, and in 2006 tuned by M.~Bargiotti~\cite{marianne}. 
\bibitem{marianne} M.~Bargiotti and V.~Vagnoni, LHCb-2007-042 (2007)
\bibitem{pdg} W.-M.Yao {\it et al.} (Particle Data Group), J.\ Phys.\ G {\bf 33}, (2006) 1 
\bibitem{zijl} T.~Sj\"ostrand and M.~van Zijl, Phys.\ Rev.\ D {\bf 36} (1987) 2019.
\bibitem{ps}T.~Sj\"ostrand and P.~Skands, Eur.\ Phys.\ J.\ C {\bf 39} (2005) 129.
\bibitem{cdfxs_jpsi} CDF Collaboration, Phys.\ Rev.\ D\ {\bf 71} (2005) 032001.
\bibitem{talkhamburg} A.~Kraan, talk given at HERA-LHC workshop, Hamburg, November 2007.
\bibitem{zongchang} Z.~Yang and S.~Qian, CMS Analysis note 2007/017 (2007).
\bibitem{hamburg}A.~Kraan, talk given at Quarkonium Workshop, Hamburg, October 2007.
\bibitem{cdfxs_ups} CDF Collaboration, Phys.\ Rev.\ Lett.\ {\bf 88} (2002) 161802.




\end{thebibliography}
\end{document}